\documentclass[epj,nopacs]{svjour}
\usepackage{graphics}

\newcommand{\la}{\left<}
\newcommand{\ra}{\right>}

\newcommand{\nvecl}{\underline{n}_l}
\newcommand{\pvec}{\ensuremath{\underline{p}}}

\newcommand{\rvec}{\ensuremath{\underline{r}}}
\newcommand{\rvecl}{\ensuremath{\underline{r}_l}}

\newcommand{\nlx}{n_{l,x}}
\newcommand{\nly}{n_{l,y}}
\newcommand{\rl}{r_l}
\newcommand{\ul}{u_l}
\newcommand{\px}{p_x}
\newcommand{\py}{p_y}
\newcommand{\pix}{p_{i,x}}
\newcommand{\piy}{p_{i,y}}
\newcommand{\rx}{r_x}
\newcommand{\ry}{r_y}

\newcommand{\kBT}{\mbox{$k_{\rm B}T$}}
\newcommand{\NPgT}{\ensuremath{\mathrm{nP}\gamma\mathrm{T}}}
\newcommand{\NVgT}{\ensuremath{\mathrm{nV}\gamma\mathrm{T}}}

\newcommand{\Tglass}{\mbox{$T_{\rm g}$}}
\newcommand{\muA}{\ensuremath{\mu_\mathrm{A}}}
\newcommand{\muAhat}{\ensuremath{\hat{\mu}_\mathrm{A}}}
\newcommand{\muF}{\ensuremath{\mu_\mathrm{F}}}
\newcommand{\muFone}{\ensuremath{\mu_1}}
\newcommand{\muFtwo}{\ensuremath{\mu_0}}
\newcommand{\Geq}{G_\mathrm{eq}}
\newcommand{\GF}{G_\mathrm{F}}
\newcommand{\Hhat}{\hat{\cal H}}
\newcommand{\Hidhat}{\hat{\cal H}^\mathrm{id}}
\newcommand{\Hexhat}{\hat{\cal H}^\mathrm{ex}}
\newcommand{\tauhat}{\ensuremath{\hat{\tau}}}
\newcommand{\tauidhat}{\hat{\tau}^\mathrm{id}}
\newcommand{\tauexhat}{\hat{\tau}^\mathrm{ex}}
\newcommand{\muhat}{\ensuremath{\hat{\mu}}}
\newcommand{\muidhat}{\hat{\mu}^\mathrm{id}}
\newcommand{\muexhat}{\hat{\mu}^\mathrm{ex}}

\newcommand{\tsamp}{\Delta t}

\newcommand{\Ubond}{U_\mathrm{bond}}
\newcommand{\lbond}{l_\mathrm{bond}}
\newcommand{\kbond}{k_\mathrm{bond}}

\newcommand{\ahat}{\hat{a}}
\newcommand{\Acal}{\mbox{$\cal A$}}

\begin{document}
\title{Numerical determination of shear stress relaxation modulus of polymer glasses}
\author{I. Kriuchevskyi 
\and J.P.~Wittmer\thanks{e-mail: joachim.wittmer@ics-cnrs.unistra.fr} 
\and O. Benzerara  
\and H. Meyer      
\and J. Baschnagel 
}                     
\institute{Institut Charles Sadron, Universit\'e de Strasbourg \& CNRS, 23 rue du Loess, 67034 Strasbourg Cedex, France}
\date{Received: date / Revised version: date}
%
\abstract{
Focusing on simulated polymer glasses well below the glass transition,
we confirm the validity and the efficiency of the recently proposed simple-average expression
$G(t) = \muA - h(t)$ for the computational determination of the shear stress relaxation modulus $G(t)$.
Here, $\muA = G(0)$ characterizes the affine shear transformation of the system at $t=0$ and 
$h(t)$ the mean-square displacement of the instantaneous shear stress as a function of time $t$. 
This relation is seen to be particulary useful for systems with quenched or 
sluggish transient shear stresses which necessarily arise below the glass transition. 
The commonly accepted relation $G(t)=c(t)$ using the shear stress auto-correlation function $c(t)$ 
becomes incorrect in this limit. 
%
} 
\maketitle
\section{Introduction}
\label{sec_intro}

\begin{figure}[t]
\centerline{\resizebox{1.0\columnwidth}{!}{\includegraphics*{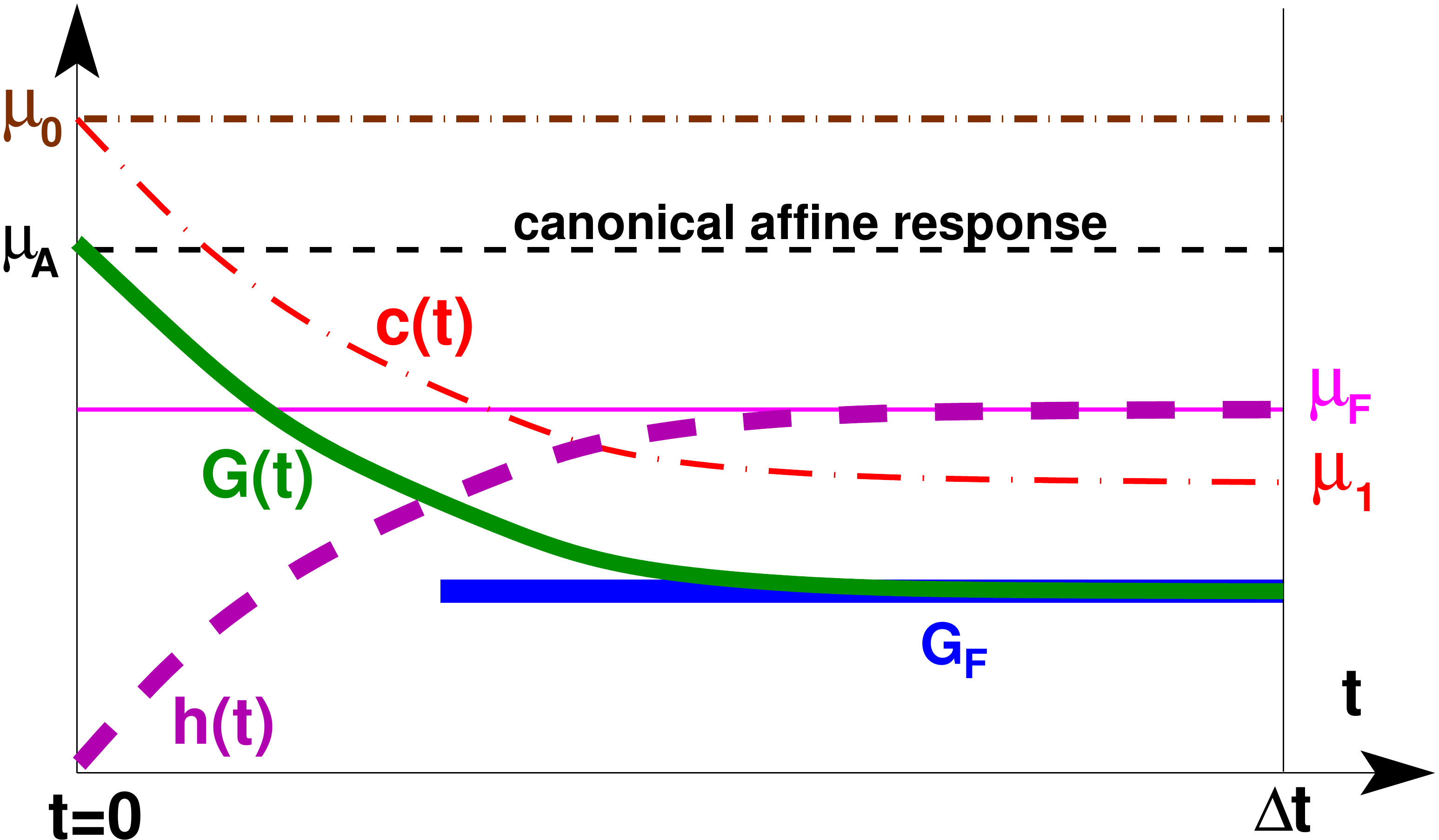}}}
\caption{Schematic sketch of several properties investigated.
The shear stress relaxation modulus $G(t)$ is indicated by the bold solid line,
the shear stress ACF $c(t)$ by the thin dash-dotted line and
the shear stress MSD $h(t)=c(0)-c(t)$ by the bold dashed line. 
Static properties are indicated by horizontal lines:
the affine shear elasticity $\muA$ (dashed line),
the shear modulus $\GF \equiv \muA - \muF$(bold solid line),
the shear stress fluctuation $\muF=\muFtwo-\muFone$ (thin solid line)
and its leading contribution $\muFtwo$ (dash-dotted line).
A canonical affine shear transformation at $t=0$ 
implies $G(t=0) = \muA$ while for large times $G(t) \to \GF$. 
At variance to this $c(t)$ decays from $c(t=0)=\muFtwo$ to $c(t=\tsamp)=\muFone$.
In general $\muA \ne \muFtwo$, hence $G(t) \ne c(t)$.
%
\label{fig_sketch}
}
\end{figure}

\paragraph*{Background.}
A central rheological property characterizing both liquids and solid elastic bodies 
is the shear relaxation modulus $G(t)$ \cite{RubinsteinBook,DoiEdwardsBook,HansenBook,AllenTildesleyBook}.
Assuming for simplicity an isotropic system, $G(t) \equiv \delta \tau(t)/\gamma$
may be obtained from the measured stress increment $\delta \tau(t) = \la \tauhat(t) - \tauhat(0^{-}) \ra$ 
after a small step strain with $|\gamma| \ll 1$ has been imposed at time $t=0$.
(As defined in Appendix~\ref{app_affine}, we denote by $\tauhat(t)$ the instantaneous shear stress 
of a configuration at time $t$.)
The direct numerical computation of $G(t)$ by means of an out-of-equilibrium simulation,
using the response to an imposed strain increment, is for technical reasons in general tedious 
\cite{AllenTildesleyBook,WXB15,WXBB15,WKB15,WXB16,WKC16}. 
It is thus of high importance to compute $G(t)$ correctly and efficiently ``on the fly" 
by means of the appropriate linear-response fluctuation-dissipation relation 
for the convenient standard $\NVgT$-ensemble at imposed particle number $n$, 
volume $V$, shear strain $\gamma$ and temperature $T$
\cite{DoiEdwardsBook,HansenBook,AllenTildesleyBook}. 
Interestingly, it is widely assumed \cite{DoiEdwardsBook,HansenBook,AllenTildesleyBook,Klix12,Szamel15} 
that quite generally 
\begin{equation}
G(t) = c(t) \equiv \la \beta V \ \overline{ \tauhat(t) \tauhat(0)} \ra
\label{eq_ct}
\end{equation}
should hold with $c(t)$ being the shear stress autocorrelation function (ACF)
and $\beta \equiv 1/\kBT$ the inverse temperature.
A bracket $\langle \ldots \rangle$ denotes here an ensemble average over $m$ independent configurations,
a horizontal bar a time average \cite{AllenTildesleyBook} for a given configuration
taken over a large, but finite sampling time $\tsamp$. 
A schematic representation of $c(t)$ is given in Fig.~\ref{fig_sketch}.
We note for later convenience that \cite{WXB16}
\begin{eqnarray}
c(t=0) & = & \muFtwo \equiv \langle \beta V \ \overline{\tauhat^2} \rangle \label{eq_muFtwo} \mbox{ and }\\
c(t=\tsamp) & = & \muFone \equiv \langle \beta V \ \overline{\tauhat}^2 \rangle. \label{eq_muFone}
\end{eqnarray}
Note that $\muFone$ does not necessarily vanish for systems with ``frozen" shear stresses 
being either permanently quenched \cite{WXB16} or transient with relaxation times much larger than the 
sampling time $\tsamp$ \cite{WKC16}. (Just as the average normal pressure $P$, a finite average shear stress $\tau$ 
may be imposed or supported by rigid walls or by the boundaries of a periodic simulation box.)
It is now well-known that the static shear modulus $\Geq$ of a given system, 
i.e. the long-time limit of $G(t)$ \cite{RubinsteinBook}, may be obtained
in the $\NVgT$-ensemble or the $\NPgT$-ensemble (at imposed average normal pressure)
using the stress-fluctuation formula 
\cite{Hoover69,Lutsko88,Barrat88,WTBL02,Barrat06,SBM11,WXP13,WXB15,WXBB15,WKB15,WXB16,WKC16}
\begin{equation}
\GF \equiv \muA - \muF \equiv (\muA-\muFtwo)+\muFone
\label{eq_GF}
\end{equation}
with $\muA$ being the ``affine shear elasticity" \cite{WXP13}, 
the Born-Lam\'e coefficient characterizing the canonical affine shear transformation of the system 
as reminded in Appendix~\ref{app_affine}, 
and $\muF \equiv \muFtwo-\muFone$ the shear stress fluctuation correcting the overestimation made by $\muA$
\cite{Lutsko88,WTBL02,Barrat06,WXBB15}.
The static properties $\muFtwo$, $\muA$, $\muF$ and $\GF$ are indicated in Fig.~\ref{fig_sketch} by horizontal 
lines.\footnote{We distinguish the material property $\Geq$ from the stress-fluctuation 
formula $\GF(\tsamp)$. $\Geq$ is more general in the sense that it does not depend on the specific 
measurement procedure, $\GF(\tsamp)$ is more general since it corresponds for stationary systems 
to a $\tsamp$-dependent moment over $G(t)$ \cite{WXP13,WXB16,WKC16},
i.e. it may characterize an intermediate plateau modulus for complex fluids 
for which the ensemble-averaged thermodynamic static modulus vanishes, 
$\Geq=\lim_{\tsamp\to \infty} \GF(\tsamp) = 0$.}
%
As emphasized in refs.~\cite{WXB15,WXBB15}, eq.~(\ref{eq_ct}) is {\em in general} not consistent with 
eq.~(\ref{eq_GF}) since for large times $c(t) \to \muFone$. This is, however, only one of the three terms 
contributing to $\GF$.\footnote{It is not helpful to consider instead the shifted ACF $C(t,\tsamp) = c(t) -\muFone(\tsamp)$ 
since, by definition, $C(t) \to 0$ for $t \to \tsamp$. Note that eq.~(\ref{eq_key})
may be rewritten as $G(t) = \GF(\tsamp) + C(t,\tsamp)$ \cite{WXB15,WXBB15}.
As pointed out in \cite{WXB16}, this formula is inconvenient
since the expectation values of both terms depend in general on $\tsamp$.}
 
\paragraph*{Tips $\&$ tricks.}
%
The first tip we want to give in the present work is that this problem is resolved quite generally 
using the more fundamental linear-response relation \cite{WKB15,WXB16,WKC16} 
\begin{equation}
G(t) = \muA- h(t) \mbox{ with } h(t) \equiv \la \frac{\beta V}{2}  \overline{(\tauhat(t)-\tauhat(0))^2} \ra
\label{eq_key}
\end{equation}
being the rescaled mean-square displacement (MSD) of the instantaneous shear stress.
This relation has been called a ``simple-average expression" in \cite{WKB15,WXB16},
since both terms $\muA$ and $h(t)$ transform as simple averages \cite{AllenTildesleyBook}
between the conjugated ensembles at constant shear strain and constant shear stress.
As a matter of fact, this is one means to derive eq.~(\ref{eq_key}) within a few lines  \cite{WKB15}.
Please note that the ACF $c(t)$ and the MSD $h(t)$ are related by \cite{DoiEdwardsBook}
\begin{equation}
h(t) = c(0) - c(t) = \muFtwo - c(t).
\label{eq_htct}
\end{equation}
Hence, $h(t) \to c(0)-c(\tsamp)= \muFtwo-\muFone= \muF$ for large times, i.e.
eq.~(\ref{eq_key}) is consistent with eq.~(\ref{eq_GF}).
Our second tip is that the already mentioned frozen shear stresses automatically drop 
out in the MSD $h(t)$. Since such frozen stresses naturally appear in  
quenched glasses, eq.~(\ref{eq_htct}) should be particular useful in this context.  
Moreover, eq.~(\ref{eq_key}) reduces to eq.~(\ref{eq_ct}) only 
if the condition
\begin{equation}
\muA=\muFtwo \mbox{ or, equivalently, } \GF =\muFone
\label{eq_condition}
\end{equation}
is satisfied. Comparing two simple static properties, 
we propose the verification of this condition as our third tip.
Please note that this condition holds, of course, for liquids:
Assuming $\tsamp$ to be larger than the longest stress relaxation time
we have $\Geq=\GF=0$ (defining property of a liquid \cite{RubinsteinBook}) 
and $\muFone=0$ (by symmetry).
Using eq.~(\ref{eq_GF}) this implies $\muA=\muFtwo$ \cite{WKC16}.

\paragraph*{Present case study.}
While in previous studies eq.~(\ref{eq_ct}) and eq.~(\ref{eq_key}) have been compared 
using permanent \cite{WXB15,WXBB15,WKB15,WXB16} or transient \cite{WKC16} elastic networks,
we focus now on slightly more realistic model systems provided by coarse-grained 
polymer glasses investigated well below the glass transition temperature $\Tglass$. 
The frozen shear stresses for each system are seen to fluctuate strongly between different configurations.
We shall show that eq.~(\ref{eq_key}) remains valid below $\Tglass$ (first tip)
and to be statistically well-behaved needing only few ($m \approx 10$) independent configurations
in the low-temperature limit despite the strong fluctuations of the frozen stresses (second tip).
It is not obvious whether eq.~(\ref{eq_condition}) also holds for polymer glasses below $\Tglass$. 
As we shall see, it does not and, consistently, eq.~(\ref{eq_ct}) is found to be incorrect (third tip).

\section{Algorithm and technical details}
\label{sec_algo}

\paragraph*{Model Hamiltonian.}
To illustrate the various points made we show below data obtained by molecular dynamics (MD) 
simulation \cite{AllenTildesleyBook} of a variant of the standard coarse-grained Kremer-Grest
bead-spring model \cite{LAMMPS}. This variant has already been used in earlier work on the
polymer glass transition \cite{SBM11,Frey15,BaschRev16}. 
It is assumed here that all monomers, that are not connected by bonds, interact via 
a Lennard-Jones (LJ) potential \cite{AllenTildesleyBook}. 
LJ units are used below.
The LJ potential is truncated at twice the potential minimum to increase numerical efficiency
and shifted there to make it continuous. 
Since it is not continuous with respect to its first derivative,
impulsive truncation corrections are required for the determination of the Born-Lam\'e coefficient 
$\muA$ as explained in \cite{XWP12}. 
The flexible bonds are represented by a harmonic spring potential
$\Ubond(r) = (\kbond/2) \ (r- \lbond)^2$
with $r$ being the distance between the connected LJ beads, 
$\kbond = 1110$ the spring constant and 
$\lbond=0.967$ the equilibrium bond length.

\paragraph*{Configuration ensemble.}
We sample dense configurations containing $M=3072$ chains of length $N=4$.
This short chain length is sufficiently long to impede the crystallization tendency
of the monodisperse LJ beads. The total number of monomer $n=12288$
is sufficient to make continuum mechanics applicable.
Periodic cubic simulation boxes are used.
The temperature $T$ and/or the normal pressure $P$ are imposed by means of the 
Nos\'e-Hoover algorithm provided by LAMMPS \cite{LAMMPS}.
A velocity-Verlet scheme with time steps of length $\delta t = 0.005$ is used.
Starting with an equilibrated configuration at $T=0.6$ we continuously cool down the 
configurations while imposing $P=0$ ($\NPgT$-ensemble). 
The average volume $V$ thus decreases slightly with decreasing temperature $T$.
This allows the determination of the glass transition temperature $\Tglass \approx 0.39$ 
by calorimetry \cite{SBM11,LXW16,BaschRev16}.
Details concerning the quench protocol may be found in \cite{SBM11}. 
We focus below on one low temperature $T=0.2$ well below the glass transition. 
After having reached this temperature we fix the volume ($\NVgT$-ensemble),\footnote{Changing 
from the $\NPgT$- to the $\NVgT$-ensemble does, of course, 
not change the average normal pressure for the large systems considered here \cite{AllenTildesleyBook}.
Concerning the stress-fluctuation formula, eq.~(\ref{eq_GF}), we could have continued with the $\NPgT$-ensemble \cite{WXP13}. 
Not possible are ensembles at imposed average shear stress $\tau$.}
temper the systems over $\tsamp=10^5$ and perform only then production runs over again $\tsamp=10^5$. 
By retempering and resampling several configurations we have verified that ageing effects
can be regarded to be irrelevant --- at least for the macroscopic properties
(averaging over the entire simulation box) of interest here.

\paragraph*{Data sampling.}
We compute time series of various instantaneous properties $\ahat$ 
with entries made each velocity-Verlet sweep over the total sampling time $\tsamp$. 
Using the expressions reminded in Appendix~\ref{app_affine} 
we write down especially the instantaneous shear stress $\tauhat$ 
and the instantaneous shear elasticity $\muAhat$ for the three shear planes $(x,y)$, 
$(x,z)$ and $(y,z)$. 
If nothing else said, we average in the end over these three equivalent shear planes
and over $m=100$ independently quenched configurations.
The average behavior and the fluctuations of an instantaneous property $\ahat$ for a given configuration 
are characterized first by computing the time-averages $\overline{\ahat}$ and $\overline{\ahat^2}$.
The expectation values are then obtained by taking the first moment $\Acal$ over the configuration ensemble.
(We thus determine, e.g., the shear stress fluctuation $\muF$.) 
To characterize also the fluctuations between different configurations 
we take in addition the second moment of the (time-preaveraged) property over the ensemble. 
We thus indicate below standard deviations $\delta \Acal$ and error bars $\delta \Acal/\sqrt{m}$.
We only consider one shear plane for the latter properties to give a conservative estimate
without any spurious correlations.

\section{Computational results}
\label{sec_res}

\begin{figure}[t]
\centerline{\resizebox{1.0\columnwidth}{!}{\includegraphics*{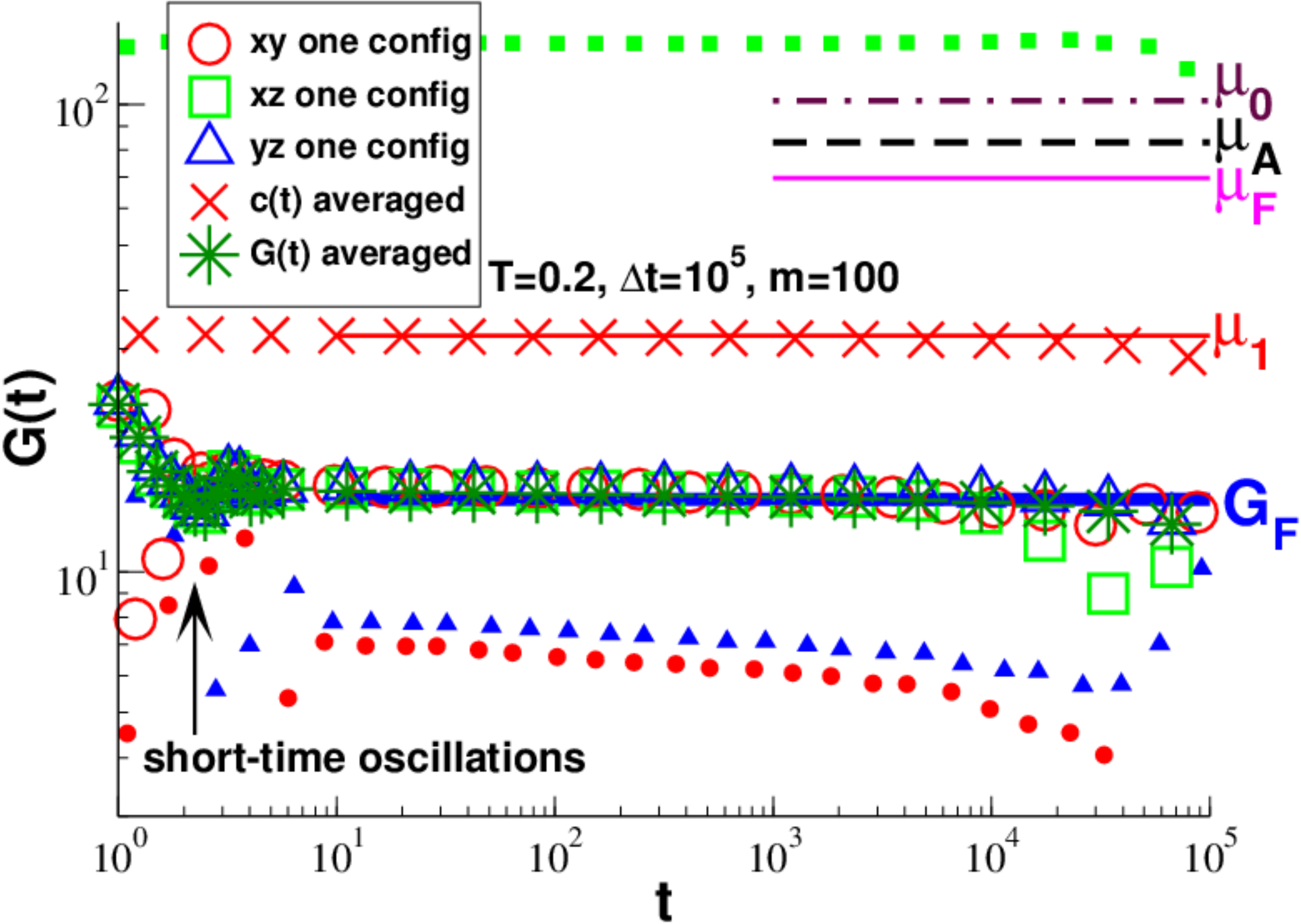}}}
\caption{Determination of $G(t)$ at $T=0.2$.
The values of the ensemble-averaged static properties $\muFtwo$, $\muA$, $\muF$, $\muFone$ and $\GF$
(from top to bottom) are represented by horizontal lines.
The filled symbols indicate for one configuration the ACF $c(t)$ for the shear planes $xy$, $xz$ and $yz$
demonstrating strong fluctuations between different shear planes.
The ensemble-averaged ACF (crosses) is similar to $\muFone$.
As indicated by the open symbols, eq.~(\ref{eq_key}) yields essentially for each shear plane
the {\em same} behaviour as the ensemble-averaged relation (stars).
\label{fig_mean}
}
\end{figure}

\paragraph*{Static properties.}
Several static and dynamical first moments $\Acal$ sampled over the configuration ensemble obtained
at our reference temperature $T=0.2$ are presented in fig.~\ref{fig_mean}. 
Let us focus first on the static properties. 
The affine shear elasticity $\muA = 83.0 \pm 0.07$ is indicated by the horizontal dashed line,
the shear modulus $\GF \approx 14.3 \pm 0.15$ determined according to the stress-fluctuation formula, 
eq.~(\ref{eq_GF}), by the bold horizontal line. The mean-squared shear stress fluctuation 
$\muF \equiv \muFtwo -\muFone$ and its two contributions are given by $\muF \approx 69.7 \pm 0.15$, 
$\muFtwo \approx 102 \pm 3.7$ and $\muFone \approx 32 \pm 3.7$.
As a consequence, 
\begin{equation}
\muFtwo - \muA \approx 19 \pm 4,
\label{eq_muFtwo_muA}
\end{equation}
i.e. the condition eq.~(\ref{eq_condition}) is clearly not satisfied. 
The errors $\delta {\cal A}/\sqrt{m}$ given above are obtained from the corresponding standard deviations $\delta {\cal A}$. 
Note that $\delta \muFtwo \approx \delta \muFone \approx 37$ 
is about the same order as the corresponding mean values $\muFtwo$ and $\muFone$.
This is nearly two orders of magnitude larger than $\delta \muF \approx \delta \GF \approx 1.5$ 
and $\delta \muA \approx 0.7$.
Since $\muF=\muFtwo-\muFone$, this implies that $\muFtwo$ and $\muFone$ must be strongly correlated, 
i.e. the dimensionless covariance coefficient of both quantities must be close to unity
(as one readily verifies directly).
The large number $m=100$ of independent configurations of this study was needed to obtain a sufficiently 
small error-bar ($\approx \delta \muFtwo/\sqrt{m}$) for the difference indicated in eq.~(\ref{eq_muFtwo_muA}).
Considering that $\muA=\muFtwo$ strictly holds in the liquid limit above $\Tglass$
(e.g., $\muA=\muFtwo \approx 75$ for $T=0.5$), 
eq.~(\ref{eq_muFtwo_muA}) is a remarkable and unexpected result.\footnote{We 
remind that $\muA = \muFtwo$ holds strictly
in systems of self-assembled transient networks created by
reversibly bridging soft spheres by harmonic springs
irrespective of the scission-recombination frequency $f$ \cite{WKC16}.
Hence, $\GF(\tsamp) = \muFone(\tsamp)$ for all $f$.
The shear modulus thus becomes finite at low $f$ due to
$\tsamp$-dependent transient shear stresses, i.e.
due to a purely dynamical effect.
}
We stress that $\muA$ and $\muFtwo$ are both proper static properties, 
i.e. their expectation values do not depend on the sampling time $\tsamp$ \cite{WXB16}.
This has been verified by comparing the averages for different sampling times $\tsamp$. 
Without entering deeper into this issue we emphasize
that in this sense around and below the glass transition some
truely static (thermodynamic) properties change with respect to the liquid state.

\paragraph*{First and second tips.}
We turn now to the dynamical properties presented in fig.~\ref{fig_mean}.
The large open symbols indicate the values of $\muA-h(t)$ obtained 
for the three different shear planes of one arbitrary configuration. 
It is seen that the data for all three shear planes are more or less identical. 
The reason for this is simply that the different frozen shear stresses of 
each shear plane automatically drop out if eq.~(\ref{eq_key}) is used. 
Interestingly, even the values of one single configuration are already
very similar to the ensemble-averaged data indicated by the large stars.
Note that $\muA-h(t)$ is similar for $t \gg 10$ to the stress-fluctuation
estimate $\GF$ of the shear modulus indicated by the bold horizontal line. 
(The thermostat is not sufficiently strong to suppress oscillations for smaller times.)
We have directly checked that the standard deviation $\delta [\muA-h(t)]$ remains below unity
for all times.\footnote{Interestingly, 
$\delta h(t) \approx \delta \muA$ for not too large times for all temperatures $T < \Tglass$.}
Assuming $m \approx 10$ configurations thus corresponds to error bars much smaller than the symbol size.
This confirms that eq.~(\ref{eq_key}) works and this with little fluctuations between different 
shear planes and configurations. 

\paragraph*{Third tip.}
The failure of the condition eq.~(\ref{eq_condition}) suggests that eq.~(\ref{eq_ct}) 
cannot be the appropriate relation for the determination of the relaxation modulus $G(t)$.
That this is indeed the case can be seen from the ACFs $c(t)$ presented in fig.~\ref{fig_mean}.
The filled symbols indicate data for the three different shear planes of the one configuration
we have already focused on above.
It is seen that the data for each shear plane is rather different and, moreover, essentially constant.
This is readily explained by noting that the ACFs are given by the (essentially)
frozen average shear stress $\overline{\tauhat}$ of each shear plane, i.e. 
\begin{equation}
\beta V \overline{\tauhat(t) \tauhat(0)} \approx \beta V \overline{\tauhat}^2
\mbox{ for } 1 \ll t \le \tsamp.
\label{eq_ct_shearplane}
\end{equation}
As shown by the crosses, $\langle \ldots \rangle$-averaging over all $m=100$ configurations and the three shear planes does not make things better. 
In agreement with eq.~(\ref{eq_muFone}), this simply leads to $c(t) \approx \muFone$ (thin horizontal line). 
This is much larger than the shear modulus $\GF$ (bold horizontal line). 
Being similar as the data presented in fig.~9 of ref.~\cite{WKC16} for self-assembled transient networks,
we find that the standard deviation $\delta c(t)$ is constant, $\delta c(t) \approx \delta \muFone$,
i.e. it is of the same order as its mean $c(t) \approx \muFone$.

\section{Conclusion}
\label{sec_conc}

\paragraph*{Summary.}

Extending our recent work on permanent and transient elastic networks 
\cite{WXB15,WXBB15,WKB15,WXB16,WKC16} to more realistic 
polymer glasses we have confirmed (validity, statistical efficiency) 
the recently proposed expression $G(t) = \muA - h(t)$ 
for the numerical determination of the shear stress relaxation modulus $G(t)$.
We have focused on one low reference temperature ($T=0.2$) in the solid limit 
and one fixed sampling time ($\tsamp=10^5$). Under these conditions plastic rearrangements 
can be neglected and strong frozen shear stresses naturally appear.
Our key relation is seen to be particularly useful under these conditions since
the frozen shear stresses --- strongly fluctuating between different shear planes and configurations ---
do directly drop out for the shear stress MSD $h(t)$.
Moreover, it was shown that the less fundamental approximation $G(t) \approx c(t)$
must fail in this limit since the condition $\muA=\muFtwo$, eq.~(\ref{eq_condition}), is violated.
As a consequence, the long-time limit $\GF$ of $G(t)$ differs from the
moment $\muFone$ of the shear stresses. The relation $G(t) \approx c(t)$
is thus not just statistically badly behaved as observed for self-assembled transient networks \cite{WKC16},
but should not be used at all.
This point constitutes a rather unexpected side-result of the presented work. 

\paragraph*{Outlook.}
While the observables presented in fig.~\ref{fig_mean} should not depend on the system size,
this is less clear for the corresponding standard deviations. For systems of self-assembled transient 
networks it can be demonstrated that $\muFtwo$, $\muFone$ and $c(t)$ reveal a strong lack of self-averaging, 
while the standard deviations $\delta \muA$, $\delta \muF$, $\delta \GF$ and $\delta [\muA-h(t)]$
decay as $1/\sqrt{V}$ \cite{WKC16,TP_V}.
It is thus likely that future simulations with larger system sizes will reveal that
the difference between the standard deviations of both sets of observables
becomes even more striking.
We note finally that we have taken advantage of the key relation eq.~(\ref{eq_key}) to 
systematically determine $G(t)$ and $\delta G(t)$ with high precision for a broad range of temperatures 
focusing especially on the behavior close to the glass transition. 
Another issue is to describe the temperature dependence of $\GF$ and $(\muFtwo-\muA)/\muA$ and 
of the reduced dimensionless standard deviations $\delta \muFone/\muFone$, $\delta \muFtwo/\muFtwo$ and $\delta \GF/\GF$.
These data will be given elsewhere.


\section*{Acknowledgments}
IK thanks the IRTG Soft Matter for financial support.
We are indebted to H.~Xu (Metz) for helpful discussions.

\appendix
\section{Canonical affine shear strains}
\label{app_affine}

Let us consider a small shear strain increment $\gamma$ in the $xy$-plane as it would be 
used to measure $G(t)$ by means of a direct out-of-equilibrium simulation
\cite{AllenTildesleyBook,WXB15,WXBB15,WKB15,WXB16,WKC16}.
For simplicity all particles are in the principal simulation box \cite{AllenTildesleyBook}.
It is assumed that all particle positions $\rvec$ and particle momenta $\pvec$
follow the imposed ``macroscopic" strain in a {\em canonical affine} manner according to \cite{WXBB15}
\begin{equation}
\rx \to \rx + \gamma \ \ry \mbox{ and } 
\px \to \px - \gamma \ \py
\label{eq_cantrans}
\end{equation}
where the negative sign in the second transform assures that Liouville's theorem \cite{Goldstein}
is satisfied.
The Hamiltonian $\Hhat$ of the configuration will thus change as
\begin{equation}
(\Hhat(\gamma)-\Hhat(\gamma=0))/V \approx \tauhat \gamma + \frac{1}{2} \muhat \gamma^2
\mbox{ for } |\gamma| \ll 1.
\label{eq_Hhatexpand}
\end{equation}
We thus define the instantaneous affine shear stress $\tauhat$ and the instantaneous 
affine shear elasticity $\muhat$ by
\begin{eqnarray}
\tauhat & \equiv & \Hhat^{\prime}(\gamma)/V|_{\gamma=0} \label{eq_tauhatdef} \mbox{ and } \\
\muhat & \equiv & 
\Hhat^{\prime\prime}(\gamma)/V|_{\gamma=0} =
\tauhat^{\prime}(\gamma)|_{\gamma=0} 
\label{eq_muhatdef}
\end{eqnarray}
where a prime denotes a functional derivative with respect to the 
imposed canonical affine transformation \cite{WXBB15}.
It follows from the last equality in eq.~(\ref{eq_muhatdef}) that
$G(t=0)=\muhat$ for the shear relaxation modulus of one configuration.
Assuming the Hamiltonian $\Hhat = \Hidhat + \Hexhat$ to be the sum of an ideal and 
an excess contribution $\Hidhat$ and $\Hexhat$, similar relations apply for the corresponding 
contributions $\tauidhat$ and $\tauexhat$ to  $\tauhat =\tauidhat + \tauexhat$ and 
for the contributions $\muidhat$ and $\muexhat$ to $\muhat = \muidhat + \muexhat$.
As shown elsewhere \cite{WXBB15} this implies for the ideal contributions
\begin{eqnarray}
\tauidhat & = & - \frac{1}{V} \sum_{i=1}^n \pix \piy / m_i \label{eq_tauidhat} \mbox{ and } \\
\muidhat & = & \frac{1}{V} \sum_{i=1}^n \piy^2 /m_i \label{eq_muAidhat} 
\end{eqnarray}
where the sums run over all $n$ particles of mass $m_i$.
Note that the minus sign for the ideal shear stress follows from the minus sign in eq.~(\ref{eq_cantrans})
required for a canonical transformation.
Assuming a pairwise central conservative potential $\Hexhat = \sum_l \ul(\rl)$
with $l$ labeling the interactions and $\rl$ the distance between the pair of monomers,
one obtains the excess contributions \cite{WXBB15}
\begin{eqnarray}
\tauexhat & = & \frac{1}{V} \sum_l \rl u^{\prime}(\rl) \ \nlx \nly   \label{eq_tauexhat} \ \mbox{ and } \\
\muexhat & = & \frac{1}{V} \sum_l  \left( \rl^2 u^{\prime\prime}(\rl)
- \rl u^{\prime}(\rl) \right) \nlx^2 \nly^2 \nonumber \\
& + & \frac{1}{V} \sum_l \rl u^{\prime}(\rl) \ \nly^2  \label{eq_muexhat}
\end{eqnarray}
with $\nvecl = \rvecl/\rl$ being the normalized distance vector.
Note that eq.~(\ref{eq_tauexhat}) is strictly identical to the
corresponding off-diagonal term of the Kirkwood stress tensor  \cite{AllenTildesleyBook}.
Similar relations are obtained for the $xz$- and the $yz$-plane.
For isotropic systems the thermal averages of all three affine shear elasticities are finite and equal.
The index ``A" indicated for historical reasons in the main text reminds that $\muA$
assumes a strictly {\em affine} strain without relaxation. It thus provides only an upper bound
$\muA = G(t=0) \ge \GF \equiv \muA- \muF \ge 0$ to the thermodynamic shear modulus
as may be seen by taking twice the derivative of the free energy with respect to $\gamma$ \cite{WXP13,WXBB15}. 
That the stress-fluctuation contribution $\muF \ge 0$ may not vanish in the zero-temperature
limit has first been emphasized by Lutsko \cite{Lutsko88}. 
See refs.~\cite{WTBL02,Barrat06,WXBB15} for details.

\bibliographystyle{epj.bst}

\end{document}